\documentclass[12pt]{article}
\usepackage{epsfig,cmb}

\def\sun{\hbox{$\odot$}}

\begin{document}

\heading{SUNYAEV-ZEL'DOVICH OBSERVATIONS WITH THE RYLE TELESCOPE}

\author{Richard Saunders} {Mullard Radio Astronomy Observatory, Cavendish
Laboratory, Madingley Road, Cambridge CB3 0HE,
UK.} {$\;$}

\begin{abstract}{\baselineskip 0.4cm 
The Ryle Telescope has been used to provide images at 15 GHz of the
Sunyaev-Zel'dovich decrements towards a dozen clusters in an X-ray
luminosity-limited sample. So far, X-ray
data have allowed $H_0$ estimates towards two of these, both giving
``low'' values that are self-consistent, though it is essential to
obtain many more estimates to reduce the effects of cluster
projection. We have also discovered a decrement towards the $z=3.8$
quasar pair PC1643+4631A\&B (198'' separation); substantial X-ray,
optical and infrared follow-up show that the cluster responsible is
either at $z>1$ or very underluminous and that very significant
gravitational lensing must be occuring.}

\end{abstract}

\section{Introduction}

The Ryle Telescope (RT) is being used to carry out a major programme
of observations of Sunyaev-Zel'dovich (S-Z) decrements \cite{s-z} towards
clusters of galaxies. This interferometer array consists of 8, 13-m
diameter dishes, currently operating at 15 GHz with a 50-K system
temperature, and a 350-MHz bandwidth correlator. Interferometers have
certain advantages over switched-beam instruments for observations of
CMB anisotropy, such as the filtering-out of much of the atmospheric
noise, the lack of need for highly stable amplifiers because only
correlated power is measured, and the fact that contaminating
radiosources can be recognised and subtracted using the longer
baselines at the same time and at the same frequency as the S-Z signal
is measured --- see e.g. \cite{saunders86,jones93,church} ad infinitum. The RT itself has two
features that make it particularly suitable for S-Z work. First, its
dishes are small: observations at cm-wavelengths are required to make
the source contamination manageable, and the angular scales of S-Z
decrements thus imply small baselines. Second, the achievement of very
long integrations without systematic offsets was a critical goal
throughout all stages of the signal-processing design and
construction.

Our main S-Z programme is targetting an unbiased sample of
ROSAT-selected clusters with X-ray luminosities $>10^{37}$ W (0.5-2.5 keV)
and with source contamination at 15 GHz of less than 5 mJy. To date we
have successful observations of a dozen clusters (see e.g. \cite{grainge93,saunders95}). I
would like to stress that there are of course firm S-Z detections
from other instruments, including the OVRO 40-m (e.g. \cite{birk}), the OVRO 5-m \cite{herb}, the OVRO interferometer \cite{carlstrom} and SUZIE \cite{willbanks}. There
is agreement in observation from a range of techniques, and the
business of S-Z observation has moved on from its checkered history to
a stage at which observation, though still challenging, is secure.

\section{Measuring $H_0$}

A key feature of S-Z astronomy is that, for a given integral of
pressure $\int n T dl$ along the line-of-sight through the cluster, the decrement
is independent of redshift $z$. Given that the X-ray surface brightness
is a line integral involving $n^2$ rather than $n$, combination of X-ray
image, S-Z image and $T$ (from X-ray spectroscopy) gives the
line-of-sight depth through the cluster. If one can turn this length into
one in the plane of the sky, then, knowing $z$ and the angular
size, one has a
determination of $H_0$. The effect of the uncertainty in the change of
view from the 90-degree turn is reduced if a suitable sample of clusters
is used; the best estimate of $H_0$ is
then the geometric mean of the estimates for each cluster. This is one
reason why our S-Z programme concentrates on clusters selected by
X-ray luminosity.

In practice, we fit a King profile for $n$ to the X-ray image to give a
best-fit gas distribution, simulate the visibilities the RT would
measure from this distribution as a function of $H_0$, and compute the
likelihood function for $H_0$ . Details of the method are given in
\cite{jones95}, which
describes its application to A2218. 

A2218 seemed to us a sensible
choice as our first target for this type of analysis. It is almost
circular in projection, so that there is reason to assume its size may
be similar along the line-of-sight. It contains no cooling flow. Indeed,
on
the scales to which the RT is sensitive, it is isothermal and is
fitted well by a King distribution. We obtained a value for $H_0$ of
$38^{+18}_{-16}$~km~s$^{-1}$~Mpc$^{-1}$. The individual errors (see below) have been combined in
quadrature.

We have applied these methods to another cluster in our sample,
A1413 at $z=0.143$. Its S-Z decrement and other properties are described
in \cite{grainge96}. It differs from A2218 in two ways. It has a cooling
flow. And it is elliptical in projection, with an axial ratio of
1.3. We are engaged in developing more-sophisticated methods for
handling such clusters, but I report here the result of what I
emphasise is our first attempt (see \cite{keiththesis}) to measure $H_0$ from A1413. The error
budget is as follows (all errors are 1-$\sigma$). There is an error in
$H_0$ of 5\% from the 3C48/3C286 calibrators (including variability),
25\% from RT noise,
10\% from source subtraction, 12\% from the ASCA estimate of $T$, and perhaps
4\% from a kinetic S-Z contribution. We allow the line-of-sight length
 to vary between the extreme projected values. Finally there is the
issue of gas clumping: the cooling flow implies clumping, and
simulations by Steve Allen (private communication) imply the gas
temperature may be underestimated by as much as 1 keV, corresponding
to an underestimate of $H_0$ by up to 10\%. With all errors combined in
 quadrature, we find from A1413 that $H_0$=$47^{+20}_{-15}$ km s$^{-1}$ Mpc$^{-1}$. It is worth noting
that both A2218 and A1413 (despite its cooling flow) have
``well-behaved'' X-ray properties on the relevant angular scales.

We have not so far been able to make $H_0$ determinations for other
clusters in our sample because of the lack of available X-ray
observations, though this situation is fortunately changing. Our best
estimate of $H_0$ is thus the geometric mean of the values for the two
clusters. It will come as no surprise to many (see \cite{adams}) that this
is 42.

\section{Finding distant clusters}

The redshift-independence of S-Z has an even more direct implication:
if they exist, one should indeed be able to detect clusters that are
too distant to be seen optically or in X-rays. Sunyaev (see
e.g.\cite{markev}) has long
emphasised that deep observation of a patch of sky should reveal an
``integrated'' S-Z effect from distant clusters. The observational
difficulty is that there are no telescopes with the requisite
combination of sensitivity and field of view to have a significant
chance of detecting a (proto-) cluster in one pointing. We therefore
have used the RT to look in the directions of three radio-quiet
quasars which, for different reasons, we judged may lie in clusters. We
found a decrement towards one of these systems, the quasar pair
PC1643+4631A\&B which have redshifts of 3.79 and 3.83 and lie 198'' apart
(see \cite{schneider}). The 6-$\sigma$ detection is shown
in Fig. 1. The signal is consistently present in different phase and
primary-beam pointings and in time-splits of the data, and checks
reveal no correlator offsets.

\begin{figure}[t!]
\begin{centering}
\centerline{\epsfig{figure=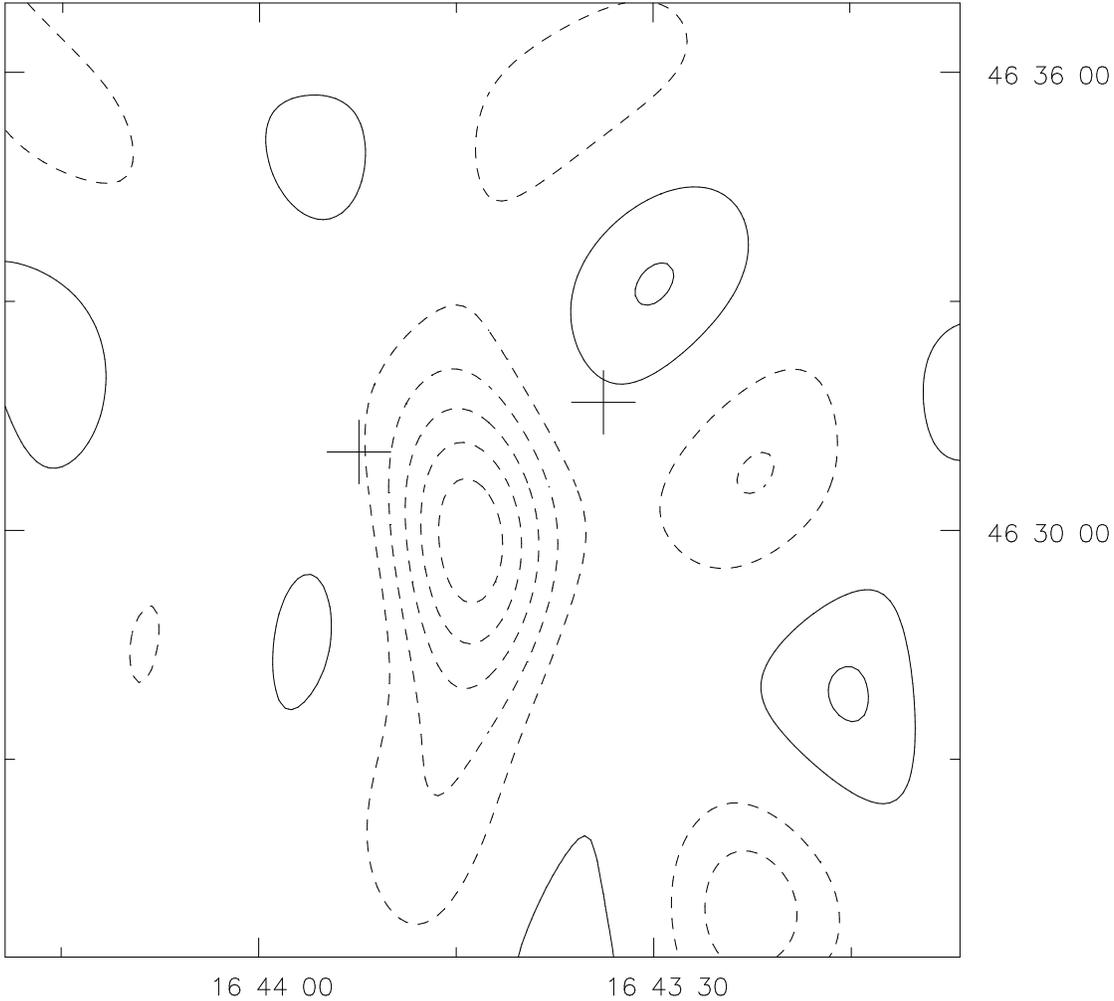,width=17cm}}
\caption{PC16433+4631: {\sc clean}ed map of the 0.65--1.25 k$\lambda$
baseline data after source subtraction. The `+' crosses indicate the
positions of quasars A (right) and B (left). Contour levels are $-325$
to +$130
\mu$Jy in steps of $65 \; \mu$Jy; dashed contours are negative.  The final map is not corrected for
primary-beam attenuation, so the noise level is uniform across the
map. Coordinates are B1950.}
\end{centering}
\end{figure}

What kind of system might produce this decrement? Such an S-Z signal
from the RT gives a {\em lower} limit to the line integral of pressure
through the (proto-) cluster. This is because the least-massive system
that can produce such an S-Z effect is one which just fills the
synthesised beam: if the cluster is smaller than the beam, there is
beam dilution and the ``true'' S-Z effect and hence cluster mass
increase; if the cluster is bigger than the beam, the cluster tends to
be resolved out and {\em again} the ``true'' S-Z effect and cluster mass
increase. Now, if the cluster has $T$ $\sim$ 5 keV (like the clusters we
know about), then, given the relation between angular size and $z$, a baryon mass of $10^{14}$ M$_{\sun}$ within a 1-Mpc radius at
$z \sim 1$ is the way to produce the S-Z effect that requires the
least mass. Given a ratio
of total mass to baryon mass of 10, the cluster mass is at least $10^{15}$~M$_{\sun}$.

The next question: is there evidence to support the notion
that the massive cluster {\it is} distant? Evidence so far comes from two
fronts. First, PC1643+4631A\&B lies on the edge of a pointed ROSAT PSPC
field; the upper limit to its X-ray flux corresponds to a luminosity
of $7 \times 10^{37}$ W (similar to the luminosities of nearby clusters in which
we see S-Z decrements) at $z>1$. Second, we have obtained deep $R$, $J$ and
$K$ images of the field with the WHT and UKIRT; there is nothing near
the decrement that looks like a cluster at $z<1$.

It thus appears that, if the cluster is like known luminous clusters, it must
lie at $z>1$ and may be an embarrassment to our standard view of
structure formation. Or it may be that the cluster is nearer, in which
case it contains far fewer luminous galaxies than known clusters, has
gas more rarefied than known clusters (to reduce the X-ray
luminosity), yet has a higher gas temperature (to maintain the S-Z).

A system of $10^{15}$ M$_{\sun}$ will gravitationally lens objects behind it. We
have carried out simple modelling of the lensing, assuming (and this
is not critical) that the lens lies at $z = 1$ and that the gas fits the
standard $\beta$-model. With a total mass of $1.2 \times 10^{15}$~M$_{\sun}$ within 1~Mpc
of the centre, the source (true) positions of quasars A and B almost
coincide, despite their observed positions being 3' apart. A tiny
adjustment to the gas distribution model would make the source
positions coincident. If A and B really are two images of one quasar,
then their spectra should be identical --- with two
exceptions. Absorption/scattering may be different along the two paths
from source to image, and the travel time is likely $\sim 10^3$ years
different so that source variability will affect the images. We have
obtained high quality optical spectra of A and B: they are remarkably
similar, though there are narrow absorption features in them that have
the same one-percent redshift difference. We are currently considering
a model in which the absorbing
gas is in the broad-line region, and that the broad-line
velocities we see have not only a Keplerian component but also a bulk
one that changes with time.

\acknowledgements{Thanks are due to Mike Jones, Jo Baker, Malcolm Bremer, Andy
Bunker, Garret Cotter, Steve Eales, Alastair Edge, Keith Grainge, Toby
Haynes, Mark Lacy, Guy Pooley and Steve Rawlings, and to the
technicians and engineers on the Ryle Telescope project, for a
superb collaborative effort. The RT is funded in part by PPARC.}

\vfill
\end{document}